\documentclass[twocolumn,english,aps,prl,floatfix,amssymb,superscriptaddress]{revtex4}
\usepackage{color}
\usepackage{amsmath}
\usepackage{amssymb}
\usepackage{graphicx}
\usepackage{esint}
\usepackage[unicode=true,pdfusetitle,
 bookmarks=true,bookmarksnumbered=false,bookmarksopen=false,
 breaklinks=false,pdfborder={0 0 1},colorlinks=true]
 {hyperref}
\hypersetup{
 dvips,linkcolor=magenta}
\usepackage{dsfont}

\makeatletter
\@ifundefined{textcolor}{}
{%
 \definecolor{BLACK}{gray}{0}
 \definecolor{WHITE}{gray}{1}
 \definecolor{RED}{rgb}{1,0,0}
 \definecolor{GREEN}{rgb}{0,1,0}
 \definecolor{BLUE}{rgb}{0,0,1}
 \definecolor{CYAN}{cmyk}{1,0,0,0}
 \definecolor{MAGENTA}{cmyk}{0,1,0,0}
 \definecolor{YELLOW}{cmyk}{0,0,1,0}
}
\ifx\pdftexversion\undefined
\else
\fi

\@ifundefined{textcolor}{}{%
 \definecolor{BLACK}{gray}{0}
 \definecolor{WHITE}{gray}{1}
 \definecolor{RED}{rgb}{1,0,0}
 \definecolor{GREEN}{rgb}{0,1,0}
 \definecolor{BLUE}{rgb}{0,0,1}
 \definecolor{CYAN}{cmyk}{1,0,0,0}
 \definecolor{MAGENTA}{cmyk}{0,1,0,0}
 \definecolor{YELLOW}{cmyk}{0,0,1,0}
}

\makeatother

\begin{document}

\title{Sign Reversal of the Hall Response in a Crystalline Superconductor}
\author{Erez Berg}
\affiliation{Department of Condensed Matter Physics, The Weizmann
Institute of Science, Rehovot, 76100, Israel}
\author{Sebastian~D.\ Huber}
\affiliation{Theoretische Physik, Wolfgang-Pauli-Strasse 27, ETH Zurich, CH-8093 Zurich, Switzerland}
\author{Netanel H. Lindner}
\affiliation{Department of Physics, Technion - Israel Institute of Technology, Haifa 32000, Israel}

\begin{abstract}
%
We consider the Hall conductivity due to the motion of a vortex in
a lattice-model of a clean superconductor, using a combination of
general arguments, unrestricted Hartree-Fock calculations, and exact
diagonalization. In the weak coupling limit, $k_F \xi \gg 1$, the sign
of the Hall response of the superconducting state is  the same as that
of the normal (non-superconducting) state.  For intermediate and strong
coupling, however, ($k_F \xi \sim 1$) we find that the sign of the Hall
response in the superconducting state can be opposite to that of the normal
state. In addition, we find that the sign reversal of the Hall response
is correlated with a discontinuous change in the density profile at the
vortex core. Implications for experiments in the cuprate superconductors
are discussed.
%
\end{abstract}

\date{\today}

\maketitle \emph{Introduction.--}The sign of the Hall response in a metal
or semiconductor reveals the charge of the underlying charge carriers
(electrons or holes). In the absence of a crystalline lattice or disorder,
the Hall conductivity is fixed by Galilean invariance, and given by
\begin{equation}
\sigma_{xy}=\frac{nec}{B},
\label{eq:ClassicalHall}
\end{equation}
where $n$ is the electron density, $c$ is the speed of light, $B$ is the
applied magnetic field, and $e$ is the the electron charge. When a lattice is
introduced, the Hall conductivity can deviate from (\ref{eq:ClassicalHall}),
both in magnitude and in sign. The explanation of the Hall conductivity
was one of the early successes of the Bloch theory of metals.

When a metal undergoes a superconducting transition, its Hall
conductivity can change dramatically. Deep in the superconducting phase,
if the superconducting vortices become pinned by disorder, the Hall
resistance vanishes (as does the diagonal resistance). If the pinning
is sufficiently weak (in clean samples, at high temperatures, or high
magnetic fields), vortex motion leads to a finite Hall resistance. If we
neglect the crystalline lattice, $\sigma_{xy}$ is again fixed by Galilean
invariance and given by (\ref{eq:ClassicalHall}); in the presence of
a lattice or impurities, however, the magnitude and sign of the Hall
response can then differ from that of the normal (non-superconducting)
state. A change in the sign of the Hall response upon approaching the
superconducting phase has been detected in the cuprate superconductors in
the underdoped \cite{Doiron-Leyraud07,LeBoeuf07,LeBoeuf11} and optimally
doped \cite{Hagen90} regime. A similar phenomenon has been observed in
conventional superconductor films \cite{Hagen90}.

The Hall response of a superconductor in the presence of
disorder has been investigated thoroughly in the literature
\cite{Otterlo95,Breznay12,Michaeli12}. The effects of the crystalline
lattice, however, have not been fully clarified. Traditionally, the problem
has been analyzed phenomenologically in a time-dependent Ginzburg-Landau
framework \cite{Dorsey92}. The key is to analyze the motion of a single
superconducting vortex in the presence of a lattice and a background
super-flow \cite{Bardeen65,Feigelman94,Otterlo95}; this is a collective
phenomena in terms of the electrons (or Cooper pairs), and is thus not
easy to describe.
\begin{figure}[b]
\includegraphics{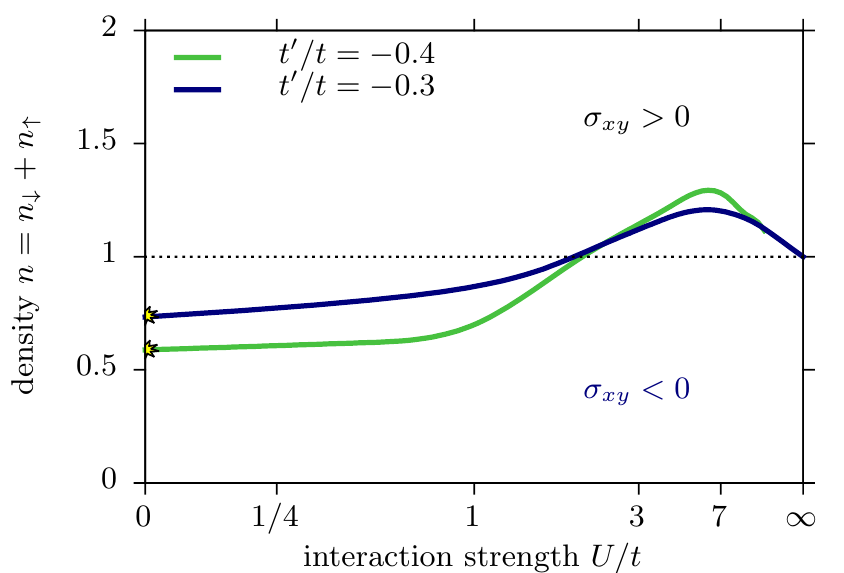}
\caption{
{\bf Phase diagram.} The Hall conductivity is given by $\sigma_{xy}
=\frac{e^2}{2\pi}( n - 2p)$ with $p\in \mathbb Z$ and $\rho$ denoting the
density. The line indicates the location where $p$ changes from zero to one
and therefore signals a sign change in $\sigma_{xy}$. The stars at $U=0$
mark the location of the van Hove singularities.
}
\label{fig:pd}
\end{figure}

In this paper, we study the Hall response of a microscopic
model of a lattice superconductor. We follow the strategy of
Refs.~\onlinecite{Lindner09,Lindner10,Huber11b}, which studied the Hall
conductivity of interacting lattice bosons: the Hall conductivity of
a system containing a single vortex is formulated as a property of the
many-body wave-function, which is calculated non-perturbatively in the
inter-particle interactions. We find a rich behavior of the Hall conductivity
as a function of the number of electrons per unit cell, $n$, and the
strength of the attractive interaction that leads to superconductivity,
$U$ (parametrized by the dimensionless number $k_{F}\xi$, where $k_{F}$
is the Fermi momentum and $\xi$ is the coherence length).

A representative phase diagram for the Hall conductivity is seen
in Fig. \ref{fig:pd}. In the Bardeen-Cooper-Schrieffer (BCS) limit,
$k_{F}\xi\gg1$, we find that the Hall conductivity changes sign as a function
of density at the point where the normal state Fermi surface changes its
topology from particle-like to hole-like. In this regime, therefore,
the sign of the Hall response is the same as in the underlying normal
state. When $k_{F}\xi$ is of the order of unity, however, the density at
which $\sigma_{xy}$ changes sign can be different from that of the normal
state. A non-monotonic behavior of this critical density as a function of
$k_{F}\xi$ is found. In a certain range of densities, the Hall conductivity
changes its sign as a function of temperature upon cooling from the normal
to the superconducting state. We discuss the origin of this behavior,
and possible implication for short-coherence length superconductors,
such as the cuprates.

\emph{Model.--}We consider a model of interacting electrons hopping on a
square lattice. The Hamiltonian is given by
\begin{align}
H & =-\sum_{\vec{r},\vec{r}',\sigma=\uparrow,\downarrow}
t_{\vec{r}\vec{r}'}^{\sigma}
c_{\vec{r}\sigma}^{\dagger}c_{\vec{r}'\sigma}^{\vphantom{\dagger}}+\mbox{H.c.}
\nonumber \\
& +\sum_{\vec{r}}\left[-U\left(n_{\vec{r}\uparrow}-\frac{1}{2}\right)
\left(n_{\vec{r}\downarrow}-\frac{1}{2}\right)-\mu n_{\vec{r}}\right].
\label{eq:H}
\end{align}
Here, $c_{\vec{r}\sigma}$ annihilates an electron on
site $\vec{r}$ with spin $\sigma$, $n_{\vec{r}\sigma} =
c_{\vec{r}\sigma}^{\dagger}c_{\vec{r}\sigma}^{\vphantom{\dagger}}$,
and $n_{\vec{r}}=n_{\vec{r}\uparrow}+n_{\vec{r}\downarrow}$. The
hopping parameters are chosen to be
$t_{\vec{r}\vec{r}'}^{\sigma}=te^{ieA_{\vec{r}\vec{r}'}^{\sigma}}$
for nearest-neighbor sites,
$t_{\vec{r}\vec{r}'}^{\sigma}=t'e^{ieA_{\vec{r}\vec{r}'}^{\sigma}}$ for
next-nearest neighbors, and $0$ otherwise. (We choose in units such that
$\hbar=c=1$.) The spin-dependent gauge field $A_{\vec{r}\vec{r}'}^{\sigma}$
is introduced in order to induce vortices in the system, and will be defined
below. $U$ is an on-site attractive interaction. When $U>0$ (which we assume
in the follwing) and $\mu\ne0$, the ground state of $H$ in Eq. (\ref{eq:H})
is superconducting for all values of $U$, crossing over smoothly from
the BCS limit $U\ll\mbox{max}\{\left|t\right|,\left|t'\right|\}$ to a
Bose-Einstein condensate of tightly bound pairs in the large $U$ limit. To
suppress a competing charge ordering instability, it will sometime
be useful to add an extended interaction term $H_{V}=V\sum_{\langle
ij\rangle}(n_{i}-1)(n_{j}-1)$, where $\langle i,j\rangle$ are
nearest-neighbor sites.

\emph{Computation of $\sigma_{xy}$.--}In order to calculate
$\sigma_{xy}$ due to the motion of a single vortex, we define the
lattice model on a torus of size $L_{x}\times L_{y}$. Next, we need
to choose the gauge field $A_{ij}^{\sigma}$. The flux of the gauge
field $A_{\vec{r}\vec{r}'}^{\sigma}$ through the system is quantized to
$2\pi N_{\phi}^{\sigma}$, where $N_{\phi}^{\sigma}$ is an integer. The
ground state is a condensate of Cooper pairs, composed of one electron
of each spin species; therefore,the total flux seen by the condensate
is $2\pi(N_{\phi}^{\uparrow}+N_{\phi}^{\downarrow})$. We see that
if $N_{\phi}^{\uparrow}=N_{\phi}^{\downarrow}$, the total number of
superconducting flux quanta is an even integer. Then, there are at least
two vortices on the torus, and the analysis of the contribution of a single
vortex to $\sigma_{xy}$ is complicated by inter-vortex interactions.

Alternatively, we may choose
$\{N_{\phi}^{\uparrow},N_{\phi}^{\downarrow}\}=\{1,0\}$, for which there
is a single vortex on the torus. This choice may look odd at first glance,
since it does not correspond to a physical magnetic field. Nevertheless,
we argue that it captures correctly the contribution of a single vortex
to $\sigma_{xy}$ in the limit $L_{x},L_{y}\gg\xi$. To understand why,
we note that the magnetic field in the vortex core region is small. As we
will argue in the following, the Hall conductivity is determined by the
structure of the core. Therefore, in the limit of large system size, the
only role of the external magnetic field is to guarantee that the ground
state has a single vortex; the dynamics of the vortex is independent of
the precise way it was induced.

In the following, we set $A_{\vec{r}\vec{r}'}^{\downarrow}=0$, and
choose $A_{\vec{r}\vec{r}'}^{\uparrow}$ such that electrons with
spin up are subject to a uniform flux of $2\pi/L_{x}L_{y}$ per unit
cell. An explicit gauge choice for $A_{\vec{r}\vec{r}'}^{\uparrow}$
is shown in \cite{SM}. We imposed twisted boundary conditions such
that the electron operators satisfy and $c_{\vec{r}+ L_{\alpha} \hat
e_\alpha,\sigma}=e^{i\Theta_{\alpha}}c_{\vec{r},\sigma}$ for $\alpha=x,y$.

The Hall conductivity of the system at $T=0$ may then be expressed as
\cite{Avron85,Fukui05}
\begin{equation}
\sigma_{xy}=\frac{e^{2}}{\left(2\pi\right)^{2}}
\int_{0}^{2\pi}\int_{0}^{2\pi}d\Theta_{x} d\Theta_{y}
\mbox{Im}
\langle\partial_{\Theta_{x}}\Psi|\partial_{\Theta_{y}}\Psi\rangle,
\label{eq:sigmaxy}
\end{equation}
where $\vert\Psi(\Theta_{x},\Theta_{y})\rangle$ is the
many-body ground state wave-function, which depends on the
boundary conditions. Eq. (\ref{eq:sigmaxy}) requires that
$\vert\Psi(\Theta_{x},\Theta_{y})\rangle$ is unique for all
$\left(\Theta_{x},\Theta_{y}\right)\in[0,2\pi]$, as is generically the
case for our finite-size system. Then, $\sigma_{xy}$ is quantized in units
of $e^{2}/2\pi$.

Eq. (\ref{eq:sigmaxy}) relates $\sigma_{xy}$ to the Berry phase accumulated
when $\Theta_{x}$, $\Theta_{y}$ are changed adiabatically. Changing
$\Theta_{x}$, $\Theta_{y}$ moves the position of the center of mass of the
vortex on the torus \cite{Huber11b}, so the Hall conductivity per site can
be viewed as the Berry's phase $\Phi_{B}$ associated with the adiabatic
motion of the vortex core around a {\em single unit cell}. This Berry's
phase is related by a ``generalized Luttinger theorem'' to the density of
charged particles in the system \cite{Paramekanti04,Huber11b,Oshikawa00}:
\begin{equation}
\Phi_{B}=2\pi\left(\frac{n}{2}+p\right),\label{eq:Phi}
\end{equation}
where $n=n_\uparrow+n_\downarrow$ is the mean number of charge
$e$ particles per unit cell, and $p$ is an integer \cite{SM}. As
$\Theta_{x,y}$ is changed by $\pi$, the vortex core moves by $L_{x,y}$
in the $x$ or $y$ direction, respectively \cite{Lindner10,Huber11b}
(note that $\Theta_{x,y}$ twists the boundary condition for both
spin flavors). Therefore, Eq. (\ref{eq:sigmaxy}) can be expressed as
$\sigma_{xy}=\frac{e^{2}}{\left(2\pi\right)^{2}}4L_{x}L_{y}\Phi_{B}$.

In the absence of a crystalline lattice, $p=0$; Eq. (\ref{eq:Phi})
then reproduces the well known result for the Berry phase
associated with the motion of a vortex in a superfluid in free
space \cite{Haldane85}. In that case, the Berry phase is directly
related to the effective Lorentz force exerted on a moving vortex:
$\vec{F}_{L}=\nu\Phi_{B}\hat{z}\times\vec{u}_{v}$, where $\nu=\pm1$
is the vorticity of the vortex and $\vec{u}_{v}$ is its velocity. In a
lattice system, however, the force on a vortex is, strictly speaking, an
ill defined concept. Nevertheless, the connection of Eq. (\ref{eq:Phi}) to
the Hall conductivity of a single vortex, through Eq. (\ref{eq:sigmaxy}),
is still valid.

Upon varying parameters of the Hamiltonian (\ref{eq:H}), the integer $p$ can
only change discontinuously via a level crossing in the many-body spectrum,
at which the integrand of Eq. (\ref{eq:sigmaxy}) is ill-defined. Since
far away from the vortex core the system is gapped \cite{comment-gap},
any level crossing must occur within the vortex core.  In the following,
we will focus our attention to the vortex core and map out the location
of the jumps of the integer p in Eq. (\ref{eq:Phi}).

\emph{General arguments.--}We are interested in $\sigma_{xy}$ as a
function of the electron density $n$ {[}tuned by the chemical potential in
Eq. (\ref{eq:H}){]}, the interaction strength $U/t$, and the next-nearest
neighbor hopping $t'/t$. Below, we discuss general arguments that can be used
to constrained the form of $\sigma_{xy}$ as a function of model parameters.

Let us first discuss the case $t'=0$. Then,
under a particle-hole transformation $\mathcal{C}$ defined through
$\mathcal{C}_{Q}c_{\vec{r},\sigma}^{\vphantom{\dagger}}\mathcal{C}_{Q}^{-1}=e^{i\vec{Q}\cdot\vec{r}}c_{\vec{r},\sigma}^{\dagger}$
where $\vec{Q}=\left(\pi,\pi\right)$, the Hamiltonian satisfies
$H\left(\mu\right)=H^{*}(-\mu)$. Because of the complex conjugation
operation, $\sigma_{xy}$ is odd under $\mathcal{C}_{Q}$:
$\sigma_{xy}\left(\mu,U/t\right)=-\sigma_{xy}(-\mu,U/t)$. Under
$\mathcal{C}_{Q}$, $n\rightarrow2-n$; therefore, $\sigma_{xy}$ must change
its sign at $n=1$ for any value of $U/t$. This change of sign occurs through
a jump in the integer $p$ in Eq. (\ref{eq:Phi}), which is associated with
a degeneracy at the vortex core.

\begin{figure}[t]
\includegraphics{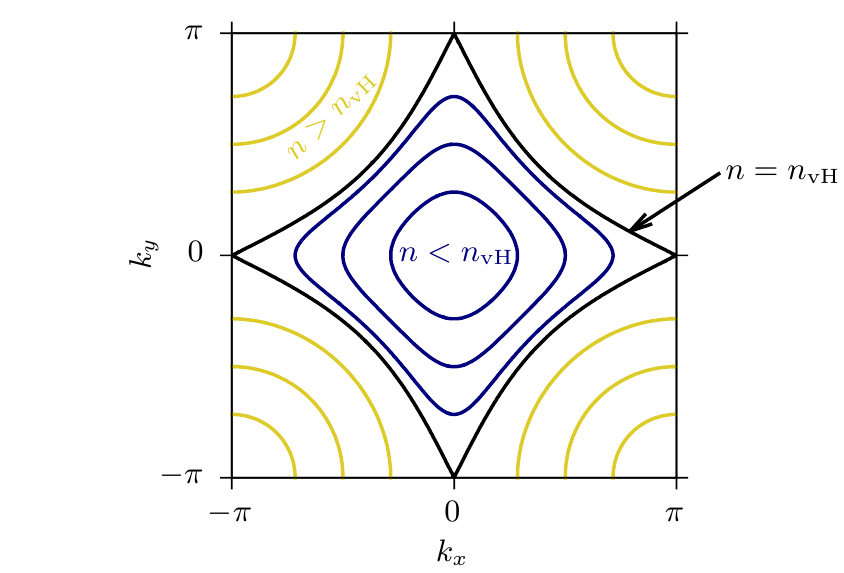}
\caption{
{\bf Fermi surface.} Fermi surface for $U=0$ and $t'/t=-0.3$ at different densities $n$, relative to the density of the van Hove singularity, $n_{\mathrm{vH}}$.\label{fig:vH}
}
\end{figure}

For $t'\ne0$, particle-hole symmetry is broken, and the critical density
$n_{c}$ at which $p$ jumps (and $\sigma_{xy}$ changes sign) can depend on
$U/t$. Nevertheless, in the extreme limits of weak and strong coupling,
the position of the jump can be deduced from the following arguments. For
$U/t=0$, the system is non-interacting with a single particle dispersion
given by $\varepsilon_{\vec{k}}=-2t\left(\cos k_{x}+\cos k_{y}\right)-4t'\cos
k_{x}\cos k_{y}-\mu$. The ground state is a filled Fermi sea. The Fermi
surface undergoes a van Hove singularity at $\mu=4t'$ (corresponding to a
density $n_{\mathrm{vH}}$), changing its character from a particle-like to a
hole-like Fermi surface (see Fig. \ref{fig:vH}). By standard semiclassical
reasoning \cite{Ashcroft87}, $\sigma_{xy}$ is expected to change sign at
$n=n_{\mathrm{vH}}$. Continuity implies that in the limit $U/t\rightarrow0$,
$n_{c}(U/t)\rightarrow n_{\mathrm{vH}}$. In \cite{SM}, we show that $n_{c}$
indeed changes continuously in the limit $U/t\rightarrow0$ and approaches
$n_{\mathrm{vH}}$, by analyzing the spectrum of a vortex core in the
weak-coupling limit.

\begin{figure}[t]
\includegraphics{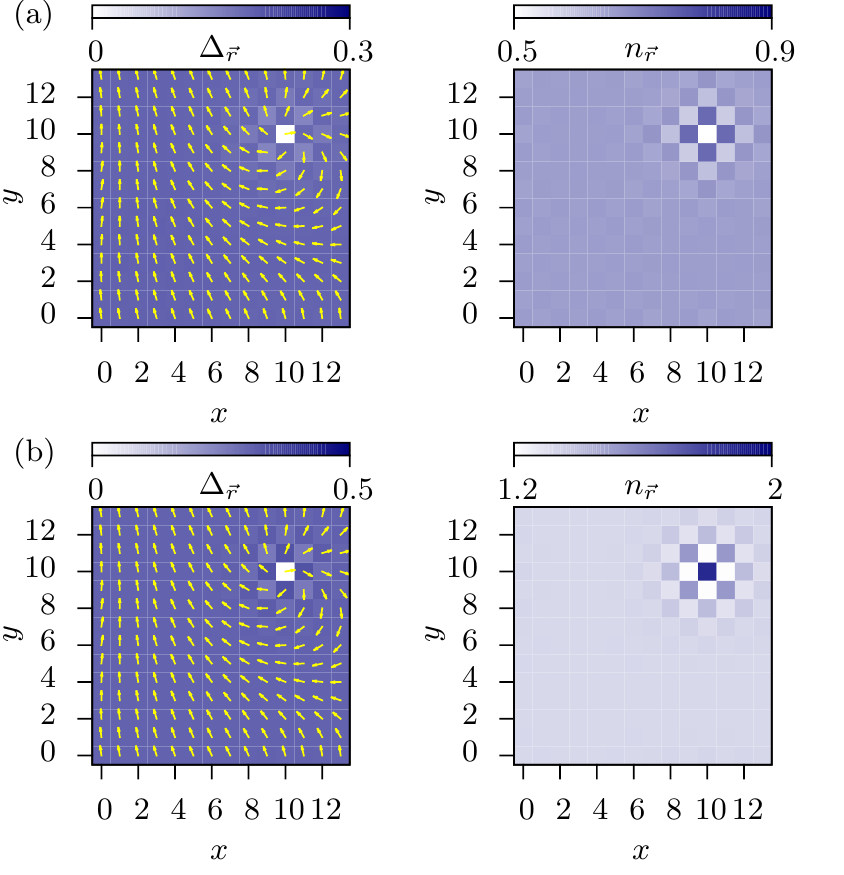}
\caption{
{\bf Vortex cores.} Order parameter (left panels) and density profiles (right panels) for $U/t=3$, $V/t=-0.05$, $t'/t=-0.3$, $L_x=L_y=L=14$. In the left panels the color indicates the amplitude of the order parameter while the arrow indicates its phase. In the right panels the color reflects the total density. (a) Particle-like superconductor at $n\approx 0.65$ where the vortex core nucleates a charge density wave with a depleted site at the vortex core. Here $\sigma_{xy}< 0$ (b) Hole-like superconductor at $n\approx 1.4$ with a charge density wave where the central site carries an excess density and $\sigma_{xy}>0$. In both cases the vortex was centered on a site by using fluxes through the openings of the torus $(\Theta_x,\Theta_y)=(10,10)\times 2\pi/L$.
}
\label{fig:vortex}
\end{figure}

In the opposite limit, $U/t\gg1$ (keeping $t'/t$ fixed), $n_{c}$ can also
be easily determined. To zeroth order in $t/U$, there are infinitely many
degenerate ground states, corresponding to an occupation of either zero
or two electrons in every site. Expanding in powers of $t/U$, one obtains
the following effective hard-core boson Hamiltonian:
\begin{align}
H_{b} & =-\sum_{\vec{r},\vec{r}'}
\tilde{t}_{\vec{r},\vec{r}'}
b_{\vec{r}}^{\dagger}b_{\vec{r}'}^{\vphantom{\dagger}}+\mbox{H.c.}
-2\mu\sum_{\vec{r}}n_{b,\vec{r}}\nonumber \\
& +\sum_{\vec{r},\vec{r}'}
\tilde{V}_{\vec{r},\vec{r}'}\left(n_{b,\vec{r}}-\frac{1}{2}\right)
\left(n_{b,\vec{r}'}-\frac{1}{2}\right)
+O\left(\frac{t^{3}}{U^{2}}\right),\label{eq:Hb}
\end{align}
where $b_{\vec{r}}^{\dagger}$ creates a pair
of electrons with opposite spins on site $\vec{r}$,
$n_{b,\vec{r}}=b_{\vec{r}}^{\dagger}b_{\vec{r}}^{\vphantom{\dagger}}$,
and $\tilde{t}_{\vec{r},\vec{r}'}$, $\tilde{V}_{\vec{r},\vec{r}'}$
are effective boson hopping and interaction, which scale as
$t^{2}/U$.  Explicit expressions for $\tilde{t}_{\vec{r},\vec{r}'}$,
$\tilde{V}_{\vec{r},\vec{r}'}$, as well as higher order terms in $t/U$,
are given in \cite{SM}.

To order $t^{2}/U$, the system is particle-hole symmetric:
$\mathcal{C}_{B}H_{b}(\mu)\mathcal{C}_{B}^{-1}=H_{b}(-\mu)$,
\emph{independently of $t'/t$}, where $C_B b_{\vec{r}}
C_B^{-1}=b_{\vec{r}}^{\dag}$. This fixes $n_{c}(U/t\rightarrow\infty)=1$,
up to corrections of higher order in $t/U$: Particle-hole symmetry breaking
terms appear in Eq.~(\ref{eq:Hb}) at order $t^{3}/U^{2}$, and their sign
depends on the sign of $t'$. In \cite{SM}, we analyze the correction to
$n_{c}(U/t)$ due to these terms, and find that for positive (negative)
$t'$, $n_{c}\left(U/t\right)$ approaches $1$ from below (above) in the
$U/t\rightarrow\infty$ limit. This is supported by our numerical calculations
which we discuss next.

\emph{Numerical calculations.--}In the intermediate interaction
regime, $U/t\sim1$, we lack a small expansion parameter and
therefore resort to a numerical calculation of $\Phi_{B}$. We
apply an unrestricted Hartree-Fock approximation by minimizing the energy
$\langle\Psi(\{\Delta_{\vec{r}},\mu_{\vec{r}\sigma}\})|H|\Psi(\{\Delta_{\vec{r}},\mu_{\vec{r}\sigma}\})\rangle$
where $\vert\Psi(\{\Delta_{\vec{r}},\mu_{\vec{r},\sigma}\})\rangle$ is
the ground state of the trial Hamiltonian
\begin{align}
H_{\mathrm{HF}} & =
-\!\!\!\!\!\!
\sum_{\vec{r},\vec{r}',\sigma=\uparrow,\downarrow}
t_{\vec{r}\vec{r}'}^{\sigma}
c_{\vec{r}\sigma}^{\dagger}c_{\vec{r}'\sigma}^{\vphantom{\dagger}}
\!+\!\mbox{H.c.}
+\sum_{\vec{r}}\Delta_{\vec{r}}^{*}
\, c_{\vec{r}\uparrow}c_{\vec{r}\downarrow}\!+\!\mbox{H.c.}\nonumber \\
 & \phantom{=}-\sum_{\vec{r},\sigma=\uparrow,\downarrow}
\mu_{\vec{r}\sigma} n_{\vec{r}\sigma}.\label{eqn:HF}
\end{align}
We determine the (self-consistent) variational parameters
$\{\Delta_{\vec{r}},\mu_{\vec{r}\sigma}\}$ via iteration. For the numerical
results presented below, we use lattices up to a size of $L_{x}\times
L_{y}=50\times50$, such that $L_{x,y}\gg\xi$.

In Fig.~\ref{fig:vortex} we show the resulting gap $\Delta_{\vec{r}}$
and density $n_{\vec{r}}$ profiles. We confirm that for the gauge
field $A_{\vec{r}\vec{r}'}^{\uparrow}$ a topological defect in the
phase field of $\Delta_{\vec{r}}$ is stabilized. The location of the
vortex is determined by the boundary conditions. For the figure we use
$(\Theta_{x},\Theta_{y})=2\pi/L\,\times(10,10)$, which in our gauge \cite{SM}
leads to a location of the vortex $\vec{r}_{\mathrm{V}}=(10,10)$.

By varying the boundary conditions $(\Theta_{x},\Theta_{y})$
we can calculate the Chern number of the many-body ground-state
$|\Psi(\{\Delta_{\vec{r}},\mu_{\vec{r}\sigma}\})\rangle$ using
Eq. (\ref{eq:sigmaxy}). Our results are in accordance with the
rule (\ref{eq:Phi}), relating $\Phi_{B}$ to the density up to an
integer. As the density is increased at a fixed value of $U/t$, the
integer $p$ changes abruptly from $0$ to $-1$ at a critical density
$n_{c}(U/t)$. Fig.~\ref{fig:pd} shows $n_{c}$ as a function of $U/t$,
for different values of the second-neighbor hopping amplitude $t'$. The
integer $p$ takes the value $0$ ($-1$) for densities below (above) $n_{c}$,
corresponding to negative (positive) $\sigma_{xy}$, respectively. As
expected, for $U/t\ll1$, $n_{c}\rightarrow n_{\mathrm{vH}}$, while for
$U/t\gg1$, $n_{c}\rightarrow1$. The critical density has a different
asymptotic behavior for $U\ll t$ and $U\gg t$: $n_{\mathrm{c}}(U/t\rightarrow
0)<1$ while $n_{c}(U/t\rightarrow\infty)>1$. Hence, we find that $n_{c}(U/t)$
has a non-monotonous dependence on $U/t$.

Fig.~\ref{fig:vortex} shows the density and pairing potential profiles for
two solutions on either side of the critical density $n_{c}(U/t)$. While
the pairing potential profiles look similar, the density profiles show a
clear distinction: Below the critical density {[}$n<n_{c}$, panel (a){]},
the vortex core is {\em depleted}, while the situation is reversed
above the critical density {[}$n>n_{c}$, panel (b){]}, where the core
carries an {\em excess density}. These two solutions cross in energy at
$n_{c}(U/t)$. Such a level crossing indicates a possible change in the Chern
number (\ref{eq:sigmaxy}) and hence the Berry phase $\Phi_{B}$. Note that the
density in the core is modulated with a wave vector $Q\sim(\pi,\pi)$. This
is a result of the competing charge density wave (CDW) instability for a
Fermi surface which exhibits some amount of nesting near half filling,
cf. Fig.~\ref{fig:vH}. In a homogeneous system, the CDW instability is
suppressed by the superconducting order. At the vortex core, however,
the vanishing of the gap $\Delta_{\vec{r}_{\mathrm{V}}}$ promotes the CDW
locally \cite{Hoffman02}.

\emph{Finite temperature.--} The zero-temperature results, summarized
in Fig.~\ref{fig:vortex}, show that there is a range of densities near
half-filling in which the sign of $\sigma_{xy}$ in the superconductor is
different from the normal state. Within this range, one expects also a
sign change in $\sigma_{xy}$ when we destroy superconductivity by raising
temperature at a fixed value of $U/t$. We now generalize our approach to
address the temperatures depedence of $\sigma_{xy}$.

\begin{figure}
\includegraphics{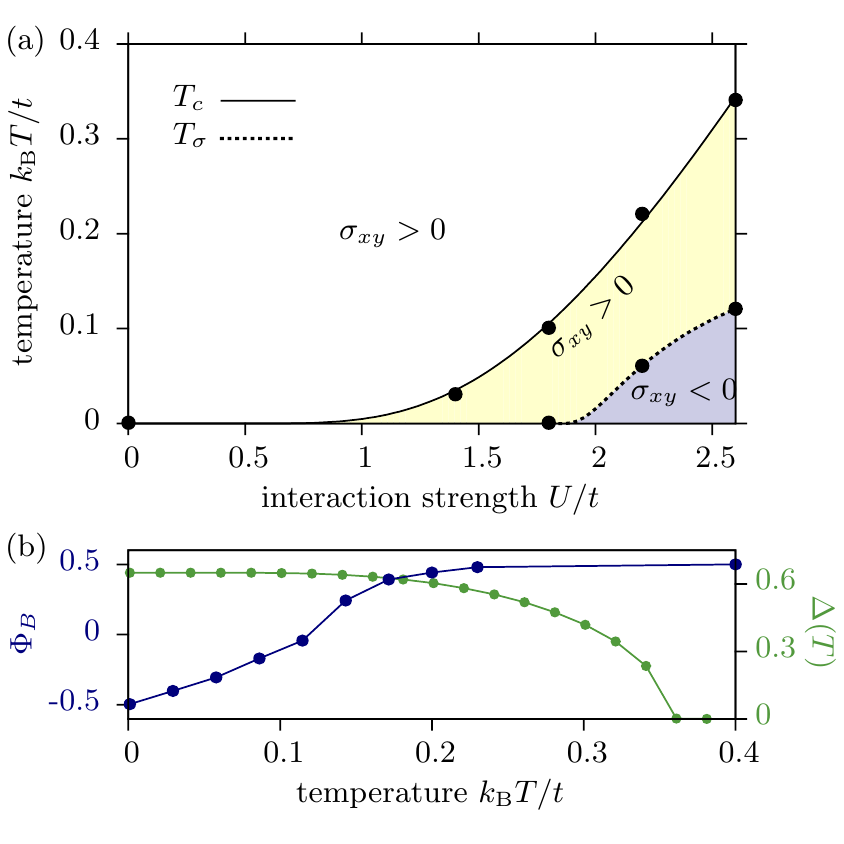}
\caption{
{\bf Finite temperature.} In panel (b) we show the dependence of the order-parameter $\Delta$ as well as the Berry phase $\Phi_B$ as a function of temperature for $U/t=2.6$, $t'/t=-0.3$, $V/t=-0.05$ and $n\approx 1$. The change in sign in $\Phi_B$ below $T_c$ is apparent and its location defines $T_\sigma$. In panel (a) we show a phase diagram for $n\approx 1$ with $T_c$ and $T_\sigma$ indicated showing that there is a sign change in $\sigma_{xy}$ {\em within} the superconducting region. Lines in panel (a) are a guide to the eye.
}
\label{fig:finite}
\end{figure}
We use a thermal state of Hamiltonian (\ref{eqn:HF})
to determine self-consistently the parameters
$\{\Delta_{\vec{r}},\mu_{\vec{r}\sigma}\}$. Then, one can calculate the
thermally averaged Chern number \cite{Lindner10}
\[
\sigma_{xy}(T)=
\sum_{\alpha}\frac{e^{2}}{\left(2\pi\right)^{2}}\int_{0}^{2\pi}
\!\!\!\!\!\!\!
d\Theta_{x}d\Theta_{y}
e^{-\frac{E_{\alpha}}{k_{\mathrm{B}}T}}
\mbox{Im}
\langle\partial_{\Theta_{x}}\!
\Psi_{\alpha}|\partial_{\Theta_{y}}\!\Psi_{\alpha}\rangle,
\]
where $\alpha$ runs over all excited states. Fig.~\ref{fig:finite} summarizes
our results for one density $n\approx1$ at $t'/t=-0.3$ and $V/t=-0.05$. In
panel (b) we show one temperature trace of $\Phi_{B}$ evaluated in a thermal
ensemble for $U/t=2.6$. We see that $\Phi_{B}$ changes its sign before
the order-parameter vanishes, i.e., within the superconductor. Moreover,
we see that the sharp jump at $T=0$ is washed out by thermally excited
states. We can now define the temperature $T_{\sigma}$ where $\Phi_{B}$
changes sign and $T_{c}$ where superconductivity is lost. Panel (a) shows
a phase diagram that summarizes our finite temperature results.

\emph{Discussion.--}We have analyzed the Hall response of a
lattice-superconductor, by examining the motion of a single vortex in
a background super-flow. By combining general arguments with numerical
simulations, we have shown that in the vicinity of half filling, it
is possible to find situations where the normal state Hall response
is \emph{opposite in sign} compared to that of the superconducting
ground state, leading to a sign change in $\sigma_{xy}$ as a function of
temperature. Unlike previously proposed mechanisms, the present mechanism
is an effect of the crystalline lattice, and does not disappear in the
clean limit.

It would be interesting to generalize our results to models applicable
to the cuprate superconductors. In particular, one needs to include the
d-wave symmetry of the order parameter and the strong correlations due to
on-site repulsive interactions. Interestingly, that the density of carriers
in the hole-doped cuprates satisfies $n_{\mathrm{vH}}<n<1$. Therefore,
within the simple model used here, it lies in the range in which the
sign of $\sigma_{xy}$ in the superconductor at intermediate coupling is
different from that of the normal state.


\emph{Acknowledgements.--} We thank A. Auerbach, G. Blatter, B. Halperin,
and S. Sachdev for useful discussions. S.H. acknowledges support from
the Swiss National Science Foundation. E. B. was supported by the Israel
Science Foundation, by the Israel-USA Binational Science Foundation,
by the Minerva Foundation, by a Marie Curie CIG grant, and by the Robert
Rees Fund. N.L. Acknowledges support from I-Core, the Israeli center of
research excellence: ``Circle of Light".

\pagebreak{} \widetext

\section*{Supplementary Material}

\setcounter{figure}{0}
\renewcommand{\thefigure}{S\arabic{figure}}

\setcounter{equation}{0}
\renewcommand{\theequation}{S\arabic{equation}}

\subsection{Choice of gauge}

\begin{figure}[b]
\includegraphics{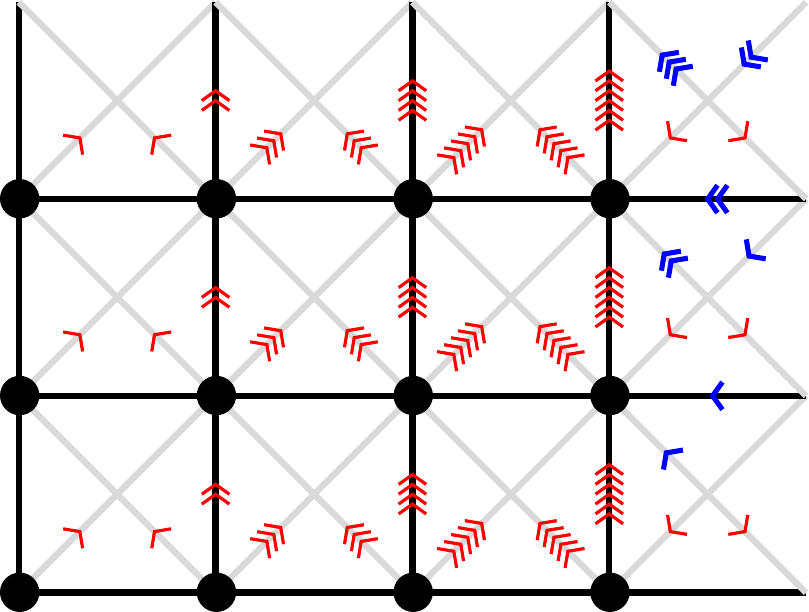}
\caption{
{\bf Gauge choice.}
}
\label{fig:Gauges}
\end{figure}
In Fig.~\ref{fig:Gauges} we show the Landau gauge we used for the
numerical calculations. The arrow indicate the phases picked up by the
$\uparrow$-electrons when hopping over the respective link. The arrows
denote the following values
\begin{equation}
\includegraphics{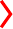} = \frac{\Phi}{2 L_x L_y} \quad \mbox{and} \quad
\includegraphics{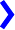} = 2L_x \times \includegraphics{single},
\end{equation}
where $L_x$ is the extent of the lattice in horizontal direction in
Fig.~\ref{fig:Gauges} and $\Phi=2\pi$ for one vortex.

Note that at the right edge of the lattice there is a winding of $2\pi$
for the hopping in the horizontal direction (blue arrows). When comparing
to the phase pattern $\phi({\vec r}) = \arg[\Delta({\vec r})]$ of the
superconducting order parameter in Fig.~\ref{fig:vortex}, we see that the
gauge-invariant current $j\sim \nabla \phi({\vec r}) - e\vec{A}({\vec r})$
is indeed continuous \cite{Lindner10}.

\subsection{Vortex Berry's phase and the Hall conductivity}
In the following we clarify the origin of the "generalized Luttinger theorem" for the Berry's phase accumulated when taking a vortex around a single unit cell.  First, consider twisting \textit{only} the boundary condition for the spin up electrons by $\Theta^\uparrow_{x,y}$, by defining the gauge field as
\begin{equation}
A^\uparrow_{\vec{r}\vec{r'}}=(A_L)^\sigma_{\vec{r}\vec{r'}}+
(\Theta^\uparrow_x,\Theta^\uparrow_y)
\cdot(\vec{r}-\vec{r'})/|\vec{r}-\vec{r'}|,
\end{equation}
where $(A_L)^\sigma_{\vec{r}\vec{r'}}$ is the gauge field configuration appearing in  Fig.~\ref{fig:Gauges}. In this case,
 changing $\Theta^\uparrow_{x,y}$ by $2\pi/ L_{x,y}$ moves the vortex by one lattice site in the $x$ or $y$ direction \cite{Lindner10, Huber11b}. Moreover, such a change in $\Theta^\uparrow_{x,y}$  and gives a Hamiltonian which is unitarily equivalent to the original one. Therefore,  a similar construction to the one appearing in Ref.~\cite{Huber11b} yields a "generalized Luttinger theorem" which gives the Berry's phase acquired when taking the vortex around a single unit cell as
\begin{equation}
\Phi_B=2\pi (n_\uparrow+p).
\label{eq: gen luttinger supp}
\end{equation}
In the model analyzed in this manuscript, $n_\uparrow=n_\downarrow=n/2$. 

Now consider twisting the boundary conditions for both spin values $\Theta^\uparrow_{x,y}=\Theta^\downarrow_{x,y}=\Theta_{x,y}$. The vortex moves by one lattice site upon changing $\Theta_{x,y}$ by $\pi/L_{x,y}$ (as a cooper pair is twisted by $2\Theta_{x,y}$). When $L_x,L_y\gg \xi$, we expect that the Berry phase accumulated when moving the vortex around a single unit cell using a twist for both spin flavours to also give $\Phi_B$ as in Eq.~(\ref{eq: gen luttinger supp}). This is indeed verified by our numerical calculations. Therefore, for $L_x,L_y\gg \xi$, we obtain the relation between the Hall conductivity and $\Phi_B$ as
\begin{equation}
\sigma_{xy}=\frac{e^2}{(2\pi)^2}L_xL_y \Phi_B.
\end{equation}
\subsection{Weak coupling limit}

For non-interacting electrons, semiclassical reasoning \cite{Ashcroft87}
shows that the sign of $\sigma_{xy}$ is determined by the topology
of the Fermi surface. For an electron-like (a hole-like) Fermi surface,
$\sigma_{xy}$ is negative (positive), respectively. In the limit
$U/t\ll1$, one may expect $\sigma_{xy}$ to approach its non-interacting
value, and hence $n_{c}(U/t)\rightarrow n_{\mbox{vH}}$ as $U/t\rightarrow0$.
This is not entirely obvious, however, since the ground state changes
its nature singularly in the limit $U\rightarrow0$. In this section,
we show explicitly that $\sigma_{xy}$ is indeed smooth for $U\rightarrow 0$.

By the arguments in the main text, the critical density $n_{c}$ in
which $\sigma_{xy}$ changes sign is associated with a level crossing
at the vortex core. Let us analyze the spectrum of Marticon-Caroli-de
Gennes (MCdG) states in the core. The Bogoliubov-de Gennes (BdG)
Hamiltonian is of the form
\begin{equation}
H_{\mbox{BdG}}=\varepsilon(\vec{k})\tau^{z}
+\Delta\left(\vec{r}\right)\tau^{+}+\mbox{H.c.}, \label{eq:HBdG}
\end{equation}
where $\varepsilon(\vec{k})$ is a general energy dispersion as a
function of the crystal momentum $\vec{k}$, $\Delta\left(\vec{r}=i\vec{\nabla}_{k}\right)$
is a pairing potential that includes a vortex at $\vec{r}=\vec{r}_{V}$,
and $\vec{\tau}$ are Pauli matrices acting in Nambu space on the
spinor $\psi_{\vec{k}}^{T} = (c_{\vec{k},\uparrow}^{\vphantom{\dagger}}, c_{-\vec{k},\downarrow}^{\dagger})$.
We assume that the system is defined on a square lattice, such that
$\varepsilon(\vec{k})$ is symmetric under $C_{4}$. The Fermi surface
is at $\varepsilon(\vec{k})=0$. For simplicity, let us consider a
linear gap function which describes a vortex at $\vec{r}_{V}=0$:
\begin{equation}
\Delta(\vec{r})=\frac{\Delta_{0}}{\xi}(x-iy),
\end{equation}
where $\Delta_{0}$ and $\xi$ are parameters. The precise gap profile
is not expected to change our results qualitatively.

\begin{figure*}[t]
\includegraphics[width=0.6\textwidth]{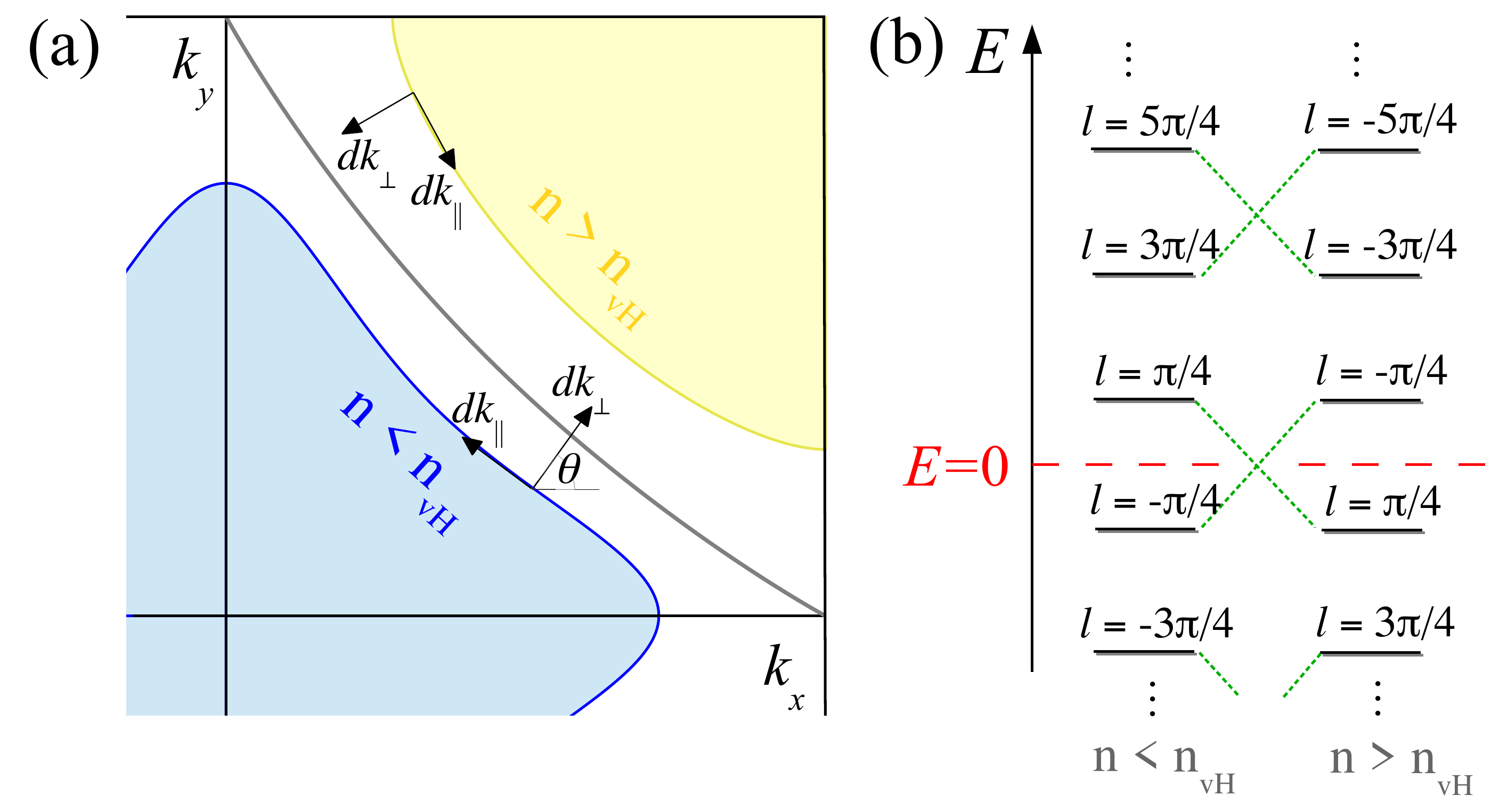}
 \caption{
(a) Transformation to the Fermi surface coordinates $(k_{\parallel}, k_{\perp})$ defined via Eq. \ref{eq:transformation}. (b) Spectrum of the BdG Hamiltonian (\ref{eq:HtBdG}), on either side of the van Hove density $n_{\mathrm{vH}}$. The states are labeled by $l$, their angular momentum under rotation by $\pi/2$ (defined mod$2\pi$). The fine dashed lines show the schematic evolution of the spectrum as $n$ is varied across $n_{\mathrm{vH}}$.
}
\label{fig:FS-weak}
\end{figure*}

We now transform the coordinates to a new frame $(k_{\parallel},k_{\perp})$
such that $k_{\parallel}$ ($k_{\perp}$) is parallel (perpendicular)
to the Fermi surface, respectively. The transformation is depicted
in Fig. \ref{fig:FS-weak}a. The new coordinates are related to the
old ones by
\begin{align}
dk_{x} & =\cos\theta dk_{\parallel}-J\sin\theta dk_{\perp},\nonumber \\
dk_{y} & =\sin\theta dk_{\parallel}+J\cos\theta dk_{\perp}.\label{eq:transformation}
\end{align}
Here, $\theta(k_{\parallel}/k_{p})$ is the angle between the normal
to the Fermi surface and the horizontal axis ($k_{p}$ is the perimeter
of the Fermi surface), and $J$ is the Jacobian of the transformation.
We fix the orientation of $k_{\parallel}$ by requiring
\begin{equation}
\int_{0}^{k_{p}}dk_{\parallel}\frac{d\theta}{dk_{\parallel}}=+2\pi.\label{eq:orient}
\end{equation}
Since Eqs. (\ref{eq:transformation}) must be total differentials,
we find that $\partial_{k_{\perp}}J=\partial_{k_{\parallel}}\theta$.
The following choice of $J$ is consistent with this constraint:
\begin{equation}
J(k_{\perp},k_{\parallel})=1+\frac{k_{\perp}}{k_{p}}\theta'\left(\frac{k_{\parallel}}{k_{p}}\right).
\end{equation}
Performing the coordinate transformation (\ref{eq:transformation}),
linearizing the dispersion near the Fermi surface, and finally performing
a similarity transformation $H_{\mbox{BdG}}\rightarrow\tilde{H}_{\mbox{BdG}}=J^{1/2}UH_{\mbox{BdG}}U^{-1}J^{-1/2}$
where $U=\exp[i\theta\tau^{z}/2]$, the BdG Hamiltonian takes the
form
\begin{equation}
\tilde{H}_{\mbox{BdG}}=H_{\parallel}+H_{\perp},\label{eq:HtBdG}
\end{equation}
where
\begin{align}
H_{\parallel} & =
\frac{\Delta_{0}}{\xi}\left(\frac{-i}{J}\right) \tau^{y}\left(\partial_{k_{\parallel}}+\frac{1}{2J}\frac{\partial J}{\partial k_{\parallel}}\right),\nonumber \\
H_{\perp} & =\frac{\Delta_{0}}{\xi} \tau^{x}i\partial_{k_{\perp}}+v_{F}(k_{\parallel})k_{\perp}\tau^{z}.
\end{align}
Here, $v_{F}(k_{\parallel})=\nabla_{k}\varepsilon(k_{\perp}=0)\cdot\hat{n}_{\perp}$
is the Fermi velocity, where $\hat{n}_{\perp}=(\cos\theta,\sin\theta)$
is unit vector normal to the Fermi surface. Note that, due to the
unitary transformation $U$, the eigenstates of (\ref{eq:HtBdG})
satisfy \emph{antiperiodic} boundary conditions as a function of $k_{\parallel}$.

Anticipating $H_{\parallel}/\Delta_{0}\sim\xi^{-1}\partial_{k_{\parallel}}\sim(k_{p}\xi)^{-1}\ll1$,
we diagonalize $\tilde{H}_{\mbox{BdG}}$ perturbatively in $H_{\parallel}$.
To zeroth order in $H_{\parallel}$, $\tilde{H}_{\mbox{BdG}}$ has
a family of zero modes parametrized by $k_{\parallel}$, of the form
\begin{align}
\varphi(k_{\perp},k_{\parallel}) 
& =\varphi_{0}[k_{\perp}a(k_{\parallel})]\left(\begin{array}{c}
1\\
-i\mbox{sgn}(v_{F})
\end{array}\right).\label{eq:phi}
\end{align}
Here, $\varphi_{0}(x)=\sqrt{2/\pi}e^{-x^{2}/2}$ is the harmonic oscillator
ground state wavefunction, and we have defined $a(k_{\parallel})=\sqrt{v_{F}(k_{\parallel})\xi/2\Delta_{0}}$.
Note that away from the van Hove point, $v_{F}\ne0$ for all $k_{\parallel}$,
and therefore $\mbox{sgn}(v_{F})$ is independent of $k_{\parallel}$.

$H_{\parallel}$ lifts the degeneracy within the zero-energy subspace
of $H_{\perp}$. Inserting $\psi(k_{\parallel},k_{\perp})=\chi(k_{\parallel})\varphi(k_{\perp},k_{\parallel})$
into the eigenvalue equation for $\tilde{H}_{BdG}$ and projecting
both sides onto the subspace of zero modes defined by Eq. (\ref{eq:phi}),
we get the following eigenvalue equation for $\chi$ {[}to leading
order in $(\xi k_{p})^{-1}${]}:
\begin{equation}
\frac{\Delta_{0}}{\xi}i\mbox{sgn}(v_{F})\left(\partial_{k_{\parallel}}-\frac{1}{a}\frac{\partial a}{\partial k_{\parallel}}\right)\chi(k_{\parallel})=E\chi(k_{\parallel}).
\end{equation}
The eigenstates are of the form $\chi_{n}=a(k_{\parallel})\exp[ik_{\parallel}/q_{n}]$,
where $1/q_{n}=\pi(2n+1)/k_{p}$, $n\in \mathds{Z}$, and the corresponding
eigenenergies are $E_{n}=-\frac{\Delta_{0}}{\xi q_{n}}\mbox{sgn}(v_{F})$.
These are nothing but the well-known MCdG states, whose minimum energy
is $\frac{\Delta_{0}}{\xi k_{p}}\sim\Delta_{0}^{2}/E_{F}$ (where
we have used the estimates $\xi\sim v_{F}/\Delta_{0}$ and $E_{F}\sim v_{F}k_{p}$).

Now, let us consider the low-energy spectrum on either side of the
van Hove point, in which the Fermi surface undergoes a change of topology.
According to our definition of the orientation of the Fermi surface,
Eq. (\ref{eq:orient}), $\mbox{sgn}(v_{F})>0$ for a particle-like
Fermi surface ($n<n_{\mbox{vH}}$), while $\mbox{sgn}(v_{F})<0$ for
a hole-like Fermi surface ($n>n_{\mbox{vH}}$). Therefore, we see
that across the van Hove point, the two states $\chi_{-1}$ and $\chi_{0}$
interchange their energy. These two states transform differently under
$C_{4}$ (e.g., under rotation by $\pi/2$, $\chi_{0}$ and $\chi_{-1}$
pick up a phases of $e^{\pm i\pi/4}$, respectively). Hence, $\chi_{0,-1}$
cannot hybridize with each other. We conclude that, as the density crosses $n_{\mathrm{vH}}$, there must be a
level crossing between $\chi_{-1}$ and $\chi_{0}$.

More generally, under rotation by $\pi/2$, $\chi_{n}$ acquires a
phase of $\exp[il_{n}]$, where $l_{n}=(2n+1)\pi/4$. Near the van
Hove point, $\chi_{2n-1}$ and $\chi_{2n}$ cross in energy. The distance
of the level crossing point to the van Hove point goes to zero in
the weak-coupling limit, $k_{p}\xi\rightarrow\infty$. The spectrum
in either side of the van Hove point is shown in Fig. \ref{fig:FS-weak}b.

In terms of the many-body spectrum, a zero energy state in the BdG
spectrum corresponds to a level crossing between the ground state
and the first excited state. Since the Chern number can only change
via a level crossing, this implies that in the weak-coupling limit,
the jump in the Hall conductivity occurs arbitrarily close to the
van Hove point. This conclusion is consistent with our numerical simulations
(Fig. 1 in the main text).

Note that our argument relies on the presence of a $C_{4}$ symmetry.
Indeed, if the $C_{4}$ symmetry is broken, there is generically no
single van Hove density in which the Fermi surface changes its character from particle-like to hole-like.
The regions of electron and hole-like Fermi surfaces are generically
separated by a density range with an open Fermi surface, in which
the normal-state Hall conductivity is ill-defined in the clean limit.

\subsection{Overlaps of Bogoliubov-de Gennes wavefunctions}

In order to calculate Chern numbers {[}Eq. (3) of the main text{]},
we need to compute overlaps of many-body wavefunctions. In the Hartree-Fock
approximation, these wavefunctions are ground states of a variational
quadratic Hamiltonian {[}Eq. (6) in the main text{]}. In order to
derive a formula for the overlap between two such wavefunctions, it
is convenient to perform a particle-hole transformation on one of
the spin species:

\begin{align}
c_{\vec{r}\uparrow} & =d_{1.\vec{r}}\nonumber \\
c_{\vec{r}\downarrow} & =d_{2,\vec{r}}^{\dagger}
\end{align}
In terms of the new operators $d_{1,2}$, $H_{\mbox{HF}}$ has the
form

\begin{align}
H_{\mathrm{HF}} & =-\!\!\!\!\!\!\sum_{\vec{r},\vec{r}',\sigma=\uparrow,\downarrow}\left(t_{\vec{r}\vec{r}'}^{\uparrow}d_{1,\vec{r}}^{\dagger}d_{1,\vec{r}'}^{\vphantom{\dagger}}-t_{\vec{r}\vec{r}'}^{\downarrow}d_{2,\vec{r}}^{\dagger}d_{2,\vec{r}'}^{\vphantom{\dagger}}+\!\mbox{H.c.}\right)-\sum_{\vec{r}}\Delta_{\vec{r}}^{*}\, d_{2,\vec{r}}^{\dagger}d_{1,\vec{r}}^{\vphantom{\dagger}}+\!\mbox{H.c.}\nonumber \\
 & \phantom{=}-\sum_{\vec{r},\sigma=\uparrow,\downarrow}\left(\mu_{\vec{r}\uparrow}d_{1,\vec{r}}^{\dagger}d_{1,\vec{r}}^{\vphantom{\dagger}}-\mu_{\vec{r}\downarrow}d_{2,\vec{r}}^{\dagger}d_{2,\vec{r}}^{\vphantom{\dagger}}\right).\label{eqn:HF-1}
\end{align}
Note that $H_{\mbox{HF}}$ contains no anomalous terms. The conservation
of the number of $d$ particles, $\hat{N}_{d}=\sum_{\vec{r}}\left(d_{1,\vec{r}}^{\dagger}d_{1,\vec{r}}^{\vphantom{\dagger}}+d_{2,\vec{r}}^{\dagger}d_{2,\vec{r}}^{\vphantom{\dagger}}\right)$,
corresponds to the conservation of the total spin in the $z$ direction
in the original problem.

One can diagonalize (\ref{eqn:HF-1}) by performing a unitary (Bogoliubov)
transformation

\begin{equation}
\left(\begin{array}{c}
\gamma_{1}\\
\vdots\\
\gamma_{N}\\
\gamma_{N+1}\\
\vdots\\
\gamma_{2N}
\end{array}\right)=U\left(\begin{array}{c}
d_{1,\vec{r}_{1}}\\
\vdots\\
d_{1,\vec{r}_{N}}\\
d_{2,\vec{r}_{1}}\\
\vdots\\
d_{2,\vec{r}_{N}}
\end{array}\right).
\end{equation}
Here, $U$ is a $2N\times2N$ unitary matrix, and $N$ is the number
of lattice points. After this transformation, $H_{\mbox{HF}}=\sum_{j=1}^{2N}E_{j}\gamma_{j}^{\dagger}\gamma_{j}^{\vphantom{\dagger}}$.
Let us order the eigenstates such that $E_{j}<0$ for $j=1,\dots,N_{0}$,
where $N_{0}$ is the number of negative energies. Then, the many-body
ground state can be written as a Slater determinant:

\begin{equation}
\vert\Psi\{\Delta_{\vec{r}},\mu_{\vec{r}\sigma}\}\rangle=\prod_{j=1}^{N_{0}}\gamma_{j}^{\dagger}\vert\tilde{0}\rangle.\label{eq:Psi}
\end{equation}
Here, $\vert\tilde{0}\rangle=\prod_{\vec{r}}c_{\vec{r}\downarrow}^{\dagger}\vert0\rangle$
is defined as the vacuum state of the $d_{1,2}$ operators, and $\vert0\rangle$
is the original vacuum of the $c_{\uparrow,\downarrow}$ operators.
Let us define $W$ as an $N_{0}\times N$ matrix containing the $N_{0}$
first rows of the matrix $U$. Then, one can verify that the overlap
between two ground states is
\begin{equation}
\langle\Psi\{\Delta_{\vec{r}},\mu_{\vec{r}\sigma}\}\vert\Psi\{\tilde{\Delta}_{\vec{r}},\tilde{\mu}_{\vec{r}\sigma}\}\rangle=\mbox{det}[W\tilde{W}^{\dagger}],
\end{equation}
where $\tilde{W}$ is the $N_{0}\times N$ matrix corresponding to
the occupied states in $\vert\Psi\{\tilde{\Delta}_{\vec{r}},\tilde{\mu}_{\vec{r}\sigma}\}\rangle$.

\subsection{Strong coupling expansion}

In the strong coupling limit, $U/t \gg 1$, the physics of the negative-$U$ Hubbard model is dominated by strongly bound pairs of electrons. We can use a Schrieffer-Wolff transformation to obtain an effective hard-core boson Hamiltonian for these pairs
\begin{align}
H_{\scriptscriptstyle \rm eff}
=&\sum_{\langle i,j\rangle} [ -J(S_i^xS_j^x+S_i^yS_j^y)+  (J+J_a)S_i^zS_j^z ]
+\sum_{\langle\langle i,j\rangle\rangle}
[ - J_2(S_i^xS_j^x+S_i^yS_j^y)+J_2 S_i^zS_j^z ]\nonumber \\
+&12J_{\rm ring}\sum_{i,jk}  S_i^zS_j^zS_k^z
-2J_{\rm ring}\sum_{i,jk} [S_i^zS_j^+S_k^- + {\rm H.c}\,].
\label{eqn:heff}
\end{align}
The different couplings can be expressed as
\begin{align}
\label{eqn:exchanges}
J&=\frac{2t^2}{U}, \quad J_2 = \frac{2t'^2}{U}, \quad J_a = 4V, \quad J_{\rm ring} = \frac{t't^2}{U^2}.
\end{align}
The first line in (\ref{eqn:heff}) contains processes up to second order in $t$ ($t'$). Note that this Hamiltonian is manifestly particle-hole symmetric at half filling {\em independently of $t'$}. The easiest way to see this is that the sub-lattice gauge-transformation $e^{i\vec{Q}\cdot\vec{r}}$ with $Q=\{\pi,\pi\}$ appearing in the particle-hole transformation for the fermions is absent in the bosonic case as no fermionic signs have to be corrected for.

The first particle-hole symmetry breaking terms appear in third order in the hopping. In (\ref{eqn:heff}) we only show those third order terms that lead to such a symmetry breaking. They all contain $J_{\rm ring}$ where exactly one hopping takes place over a next-to-nearest neighbor bond. Note that we suppress the gauge field in (\ref{eqn:heff}) for simplicity.

\begin{figure*}[t]
\includegraphics{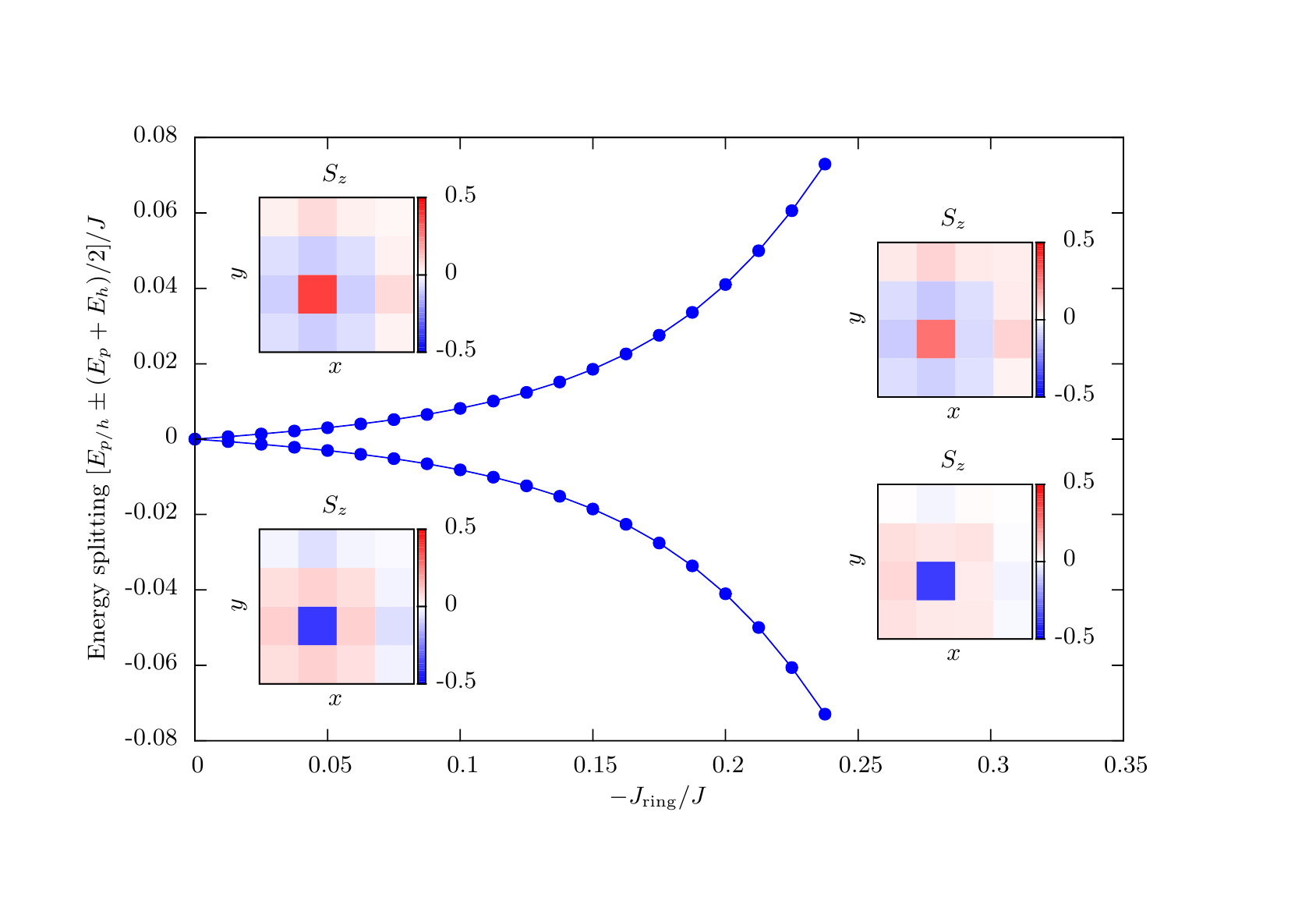}
\caption{
{\bf Strong coupling.} Energy of the two lowest states at half filling in the hard-core boson model. At $t'=0$ there is a degeneracy between two different vortices. At non-zero values of $J_{\rm ring}$ the particle-type condensate with a vortex with a density-depletion forms in the ground state.
}
\label{fig:strong}
\end{figure*}

To study the effect of the particle-hole symmetry breaking terms $\propto t^2t'$ we use exact diagonalization on clusters up to $L_x \times L_y = 4 \times 4$ using the hard-core boson model. We investigate the half-filled lattice and observe a degeneracy between two different vortex states at $t'=0$ as expected, cf. Fig.~\ref{fig:strong}. When turning on a $t'/t < 0$ as in the main text we find that the vortex with a density depletion is lower in energy. From this we conclude that the line where the integer $p$ jumps from zero to $-1$ moves towards densities $n>1$, in accordance with the results in the main text. These findings are also supported by the calculation of the Chern number \cite{Huber11b}. For the actual calculations we used values for $J_2$, $J_a$, and $J_{\rm ring}$ that don't obey the rules (\ref{eqn:exchanges}) for the following reason: Our aim is to study the breaking of particle-hole (PH) symmetry by the application of $J_{\rm ring}$ terms. For that we need a $t'$ in the fermionic Hamiltonian. However, this induces also $J_2$, which does not break PH symmetry but which has to be counter-acted by a larger $J_a$ in order to fight the competing CDW. To keep the vortex smaller than the system size, we use $J=1$, $J_2=0.2$, $J_a=0.7$ and we change $J_{\rm ring}$ from 0 to 0.2. There is no reason to expect that by changing $J_2$ and $J_a$ the main conclusion would change.

\begin{thebibliography}{22}
\bibitem[{com()}]{comment-gap}
\bibinfo{note}{In this statement, we have ignored the Goldstone modes in the
  superfluid phase, whose gap goes to zero as the system size increases.
  However, this gap is not expected to vary significantly as a function of
  $\Theta_x$, $\Theta_y$. Moreover, in a charged superconductor, there are no
  gapless Goldstone modes. One can imagine adding to the Hamiltonian
  (\ref{eq:H}) a repulsive long-range interaction which decays logarithmically
  with distance. This interaction gaps out the Goldstone modes. However, if the
  coefficient of the long-range interaction is sufficiently small, it does not
  affect the spectrum of the core much, and therefore it does not shift the
  position of the level crossing that occurs at the core.}
\end{thebibliography}
\end{document}